\documentclass[12pt]{article}

\usepackage{graphicx}
\usepackage{scicite}
\usepackage{amsmath,amssymb,amsthm}
\usepackage{times}
\usepackage{epstopdf}

\title{Integral Control Feedback Circuit for the Reactivation of Malfunctioning p53 Pathway.}

\author
{Hsu Kiang Ooi,$^{1\ast}$  Lan Ma$^{1}$\\
\\
\normalsize{$^{1}$Department of Bioengineering, The University of Texas at Dallas.}\\
\\
\normalsize{$^\ast$To whom correspondence should be addressed; E-mail:  hsukiang@utdallas.edu.}
}
\date{}

\begin{document} 


\baselineskip24pt


\maketitle

\section*{Introduction}

Many biological systems inherently possess the natural ability to adapt to environmental changes in order to maintain survivability. This ability of a biological system's sensory adaptation give rise to homeostasis. There has been numerous report of homeostasis of biological systems such as heat shock proteins, glucose level and bile acid level regulations\cite{mallouk1999,tovey1988,kim2011,armata2010}. In responding to external perturbations, homeostasis refers to the ability to sense and regulate its internal state in order to remain in a healthy state. It is also important to note here that the sensory adaptation responds to a temporary change in the external signals but eventually return to its pre-stimulus state when the perturbations diminishes. Perfect adaptation to step input perturbation is a critical feature observed in homeostasis. 

Integral feedback control is recognized to be a necessary scheme to achieve \textit{perfect adaptation}. In the context of biological systems, integral control has been identified biological systems such as bacterial chemotaxis \cite{yi2000,csete2002,elsamad2002,baker2006,hamadeh2011} and reported in engineered synthetic biology system such as a two promoter network \cite{ang2010}. The robustness in the feedback structure relies on the integral control or in general the action of integral control is implied by the Internal Model Principle (IMP) \cite{francis1976,sontag2003}. The internal model principle states that for a system, \textit{P}, it is necessary to internally encompass a subsystem, \textit{$P_{IM}$} which is able to generate control variables in response to disturbances. When the disturbance rejection or internal stability is solved, an exact duplicate of the inverse of the denominator of the disturbance signal must be present in the feedback loop at the summation junction where the original disturbance is subjected. It has been long established, for the case of step input perturbations, the adaptation can be perfect and exhibit robustness when the integral feedback control is present \cite{francis1976,sontag2003}. Step input perturbation refers to the input signal that remains present and constant-valued after it is applied.

In a previous work, the theory of integral feedback control was applied to a generic two-promoter genetic regulatory network and the constraints to design an integral feedback controller that achieve perfect adaptation was discussed in detail \cite{ang2010,ang2013}. The mathematical modeling framework proposed, guides the design of an integral feedback controller by using an \textit{Escherichia coli} two-promoter network and it shows that it is possible to tune parameters in the model to optimize the network performance in response to step input perturbation \cite{ang2010,ang2013}. We apply the idea of this integral control framework to construct a simple transcription activator like effectors (TALEs) based controller that can restore the function of p53 in diseased phenotype. TALEs can drive activation and repression of target genes in cells by fusing of transactivation domains to TALEs DNA binding domains \cite{li2012}. TALEs are discovered in \textit{Xanthomonas} bacteria which modulate gene expression in host plants and to facilitate bacterial colonization and survival \cite{boch2010}.

The core module of the p53-MDM2 network is adapted from the Ma et al. \cite{ma2005}. We expand this established model to include miRNAs regulator of p53 and MDM2. Studies suggest that microRNAs directly and/or indirectly regulate the biochemical process of p53-induced apoptosis \cite{hanahan2011,lujambio2012}. We design integral feedback controller to achieve perfect adaptation of a p53-MDM2-miRNA network under stepwise perturbation of p53. The simulation and analysis elucidate critical TALEs parameters that are tunable in an experimental setting to guide the design of optimized integral control feedback (ICF) synthetic circuits to maintain p53 basal level. The constraints imposed upon the choice of parameters to achieve a desired p53 state is also analyzed.

\begin{figure}
\begin{center}
\includegraphics[width=1\textwidth]{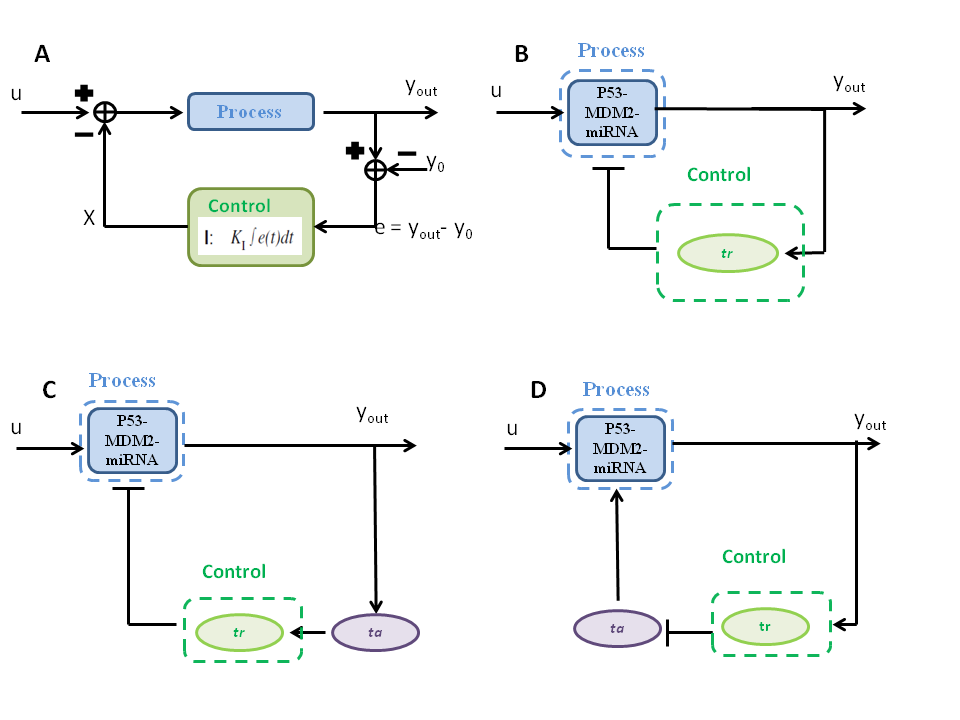}
\end{center}
\caption{(A) Block diagram showing the basic control theory of an integral control. The process controlled by linear integral feedback. The control, operates linearly on the output error, e, by direct integration. U represents the stepwise perturbation. (B) ICF Architecture 1: Schematic diagram of the ICF circuit where TALE represses p53 to complete the feedback loop. TALE respond to perturbation through its control action. (C) ICF Architecture 2: An additional TALE activator, \textit{ta} is added to the network to transcriptionally activate TALE repressor, \textit{tr}. (D) ICF Architecture 3: The TALE repressor, \textit{tr} represses a TALE activator, \textit{ta} which transcriptionally activates p53.}
\label{fig:icf1}
\end{figure}

The linear control system theory is shown in Figure \ref{fig:icf1}(A). In the block diagram, the difference between the system output, $y_{out}$ and setpoint value, $y_0$ give rise to the steady state error, $e$ of the system. Perfect adaptation is achieved when the integral control consistently eliminate the steady state error. The integral control of a system shown in Figure \ref{fig:icf1}(A) ensure perfect adaptation in the presence of step disturbances. In control theory, the integral control operates linearly on the steady state output error, $e$ given by the equation:
\begin{align}
x &= K_I \int e \ dt\\ 
\frac{d[x]}{dt} &= K_I \ e(t)\\
\frac{d[x]}{dt} &= K_I \ (y_{out} - y_0).
\end{align}
\noindent where $K_I$ is the integral gain, $X$ is the output of the control, $y_{out}$ is the steady state output and $y_0$ is the set-point. The control eliminates the steady state output error as time goes to $\infty$, the output of the process, $y_{out}$ approaches setpoint value, $y_0$. The steady-state error is eliminated as time goes to $\infty$ and the system is adapting.

While the integral control theory described above is from a linear system perspective, the p53-MDM2-miRNA network and the ICF synthetic circuits are nonlinear systems. In particular, the ICF synthetic circuits are described by nonlinear Michaelis-menten kinetics equations. For generalization, the ICF synthetic circuit (TALE) is represented by a multidimensional controller while the p53-MDM2-miRNA network is represented by the multidimensional process. The basic process and controller elements can be represented by first-order time-invariant ordinary differential equations,
\begin{align}
\frac{dy}{dt} &= g(u,x,y)\\
\frac{dx}{dt} &= f(y).
\end{align}
Perfect adaptation is achieved only when the system steady-state output value, ($y$) is independent of its input value, ($u$). Therefore, the function $f(y)$ is independent of input, ($u$) and control action, ($x$). The control action, ($x$) only depends on the system output, ($y$). For the case where the function $f(y)=0$ has a single root which is unique and independent of the input, ($u$), the corresponding output value, ($y$) is the value at which the controller will achieve at steady state. The steady state of the overall process-controller system is defined by this output value, ($y$). 

The p53-MDM2-miRNA network is represented by a multidimensional process, $y_n$ which takes the scalar input signals $u$ and control variable $x$. This multidimensional process generates the scalar output signal $y_n$ as described by 
\begin{align}
\frac{dy_1}{dt} &= g_1(u,x,y_1,... y_n)\\
\vdots \\
\frac{dy_n}{dt} &= g_n(u,x,y_1,... y_n),
\end{align}
where $n$ is the dimensionality of the process and the process output $y$ is represented by $y_n$.

The ICF synthetic circuit can be represented by a multidimensional controller. The multidimensional controller takes $y_n$ as the scalar input signal and generates the scalar-valued control action, $x_m$ described by
\begin{align}
\frac{dx_1}{dt} &= f_1(x_1,... x_m,y)\\
\vdots \\
\frac{dx_m}{dt} &= f_m(x_1,... x_m,y),
\end{align}
where $m$ is the dimensionality of the controller and $x_m$ is the controller's output signal. The multidimensional controller for the ICF circuit contains an integrator as shown in Equation \ref{eqn:integrator}:
\begin{align}
\frac{dx_1}{dt} &= f_1(x_1,... x_{i-1},y)\\
\vdots \\
\frac{dx_{i-1}}{dt} &= f_{i-1}(x_1,... x_{i-1},y)\\
\frac{dx_{i}}{dt} &= f_{i}(x_{i-1}) \label{eqn:integrator}\\
\frac{dx_{i+1}}{dt} &= f_{i+1}(x_i,... x_{m})\\
\vdots \\
\frac{dx_{m}}{dt} &= f_{m}(x_1,... x_{m})
\end{align}
The definition of an integral control in the multidimensional controller mandates that signal from the process converge and pass through this integrator (i.e. Equation \ref{eqn:integrator}) without any internal loops that may bypass the integrator. Specifically, the input signal to the controller, $y$, passes through $i-1$ coupled sub-components, eventually converging into a single signal, $x_{i-1}$, before integration. The integrated signal, $x_i$, then passes through an additional $m-i$ coupled sub-components before exiting the controller as $x_m$, the control action signal.

The requirement of perfect adaptation states that the system must have a single steady state and $f_i$ has a single root. We can then determine the steady state of the controller by setting the time derivatives of $x_1$ till $x_m$ to 0. This means that for the function $f_i(x_{i-1,ss}) = 0$, the control variable $x_{i-1,ss}$ is constant which implies $x_{i-1}$ is perfectly adapting. The control variable constant $x_{i-1,ss}$ when substituted into the first $i-1$ steady-state equations gives us a system that consist of the equations $f_1 = f_2 = ... = f_{i-1} = 0$ and variables $ x_{1,ss}, x_{2,ss}, ..., x_{i-2,ss},$ and $y_{ss}$. These variables are constants (system has a single steady state), therefore $x1, x2, ..., x_{i-2},$ and $y$ are also perfectly adapting.

\section{Results}
\label{s:res4}
In Figure \ref{fig:icf1}(A), a block diagram for an integral control feedback system consists of the process and the controller. The process is subjected to an input signal, $u$, and results in an output signal, $y_{out}$. This output signal is then fed as an input to the controller. As a result, the controller generates a control action signal, $x$. This control action signal is fed back to the process.

We designed ICF circuits to achieve adaptation of p53 when the basal level of p53 is inhibited to a constant low level (inhibition equivalent to a stepwise repression). In Figure \ref{fig:icf1}(B), the process is the p53-MDM2-miRNA network while the controller is the synthetic circuit, $tr$ (TALE). This architecture implements the integral control feedback shown in Figure \ref{fig:icf1}(A). We extend this architecture, where additional component, $ta$ is integrated into the controller before and after the $tr$, as shown in Figure \ref{fig:icf1}(C) and Figure \ref{fig:icf1}(D) respectively.

\begin{figure}
\begin{center}
\includegraphics[width=1\textwidth]{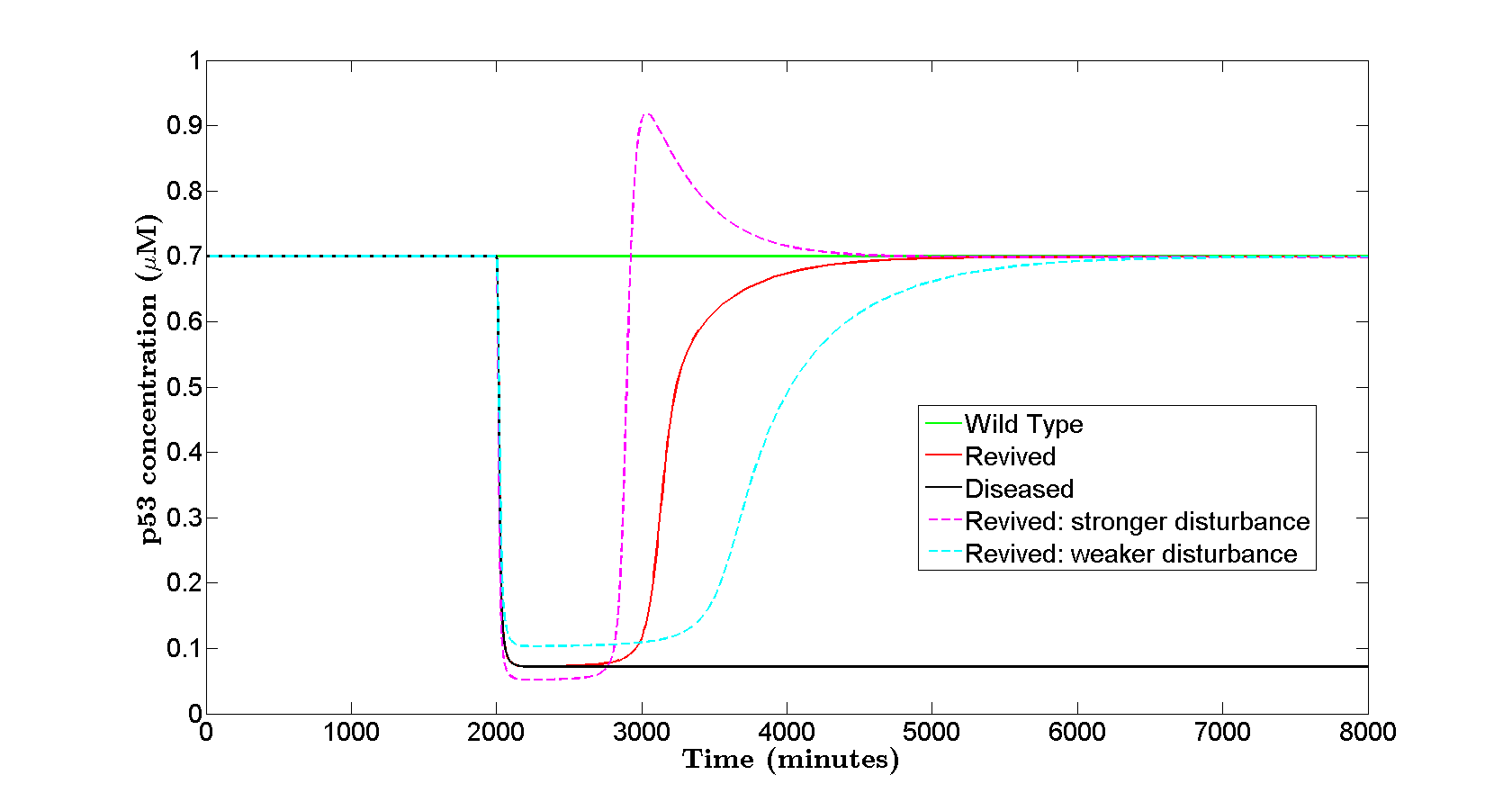}
\end{center}
\caption{Time trajectory of total p53 (active (p53*) and inactive p53 proteins) concentration of the p53-miRNA model coupled with ICF architecture 1 circuit. The wild-type (green) shows the typical concentration level of total p53 in a healthy cell. The diseased state (black) shows are critically low concentration level of p53 after input perturbation introduced at t=2000 minutes without the ICF architecture 1 circuit. When the ICF (red, $u=0.66$) is introduced into the model, the p53 level return to its wild-type state. Two other scenario is also simulated; a weaker (cyan, $u=0.73$) and stronger (magenta, $u=0.60$) perturbation strength was also introduced. The p53 concentration levels show adaptation where the weaker perturbation shows delayed response and the stronger perturbation shows transient response.}
\label{fig:trajicf1}
\end{figure}

\subsection{Architecture 1: Single Stage Integral Control Feedback Circuit.}
To achieve perfect adaptation, it is necessary to maintain a negative feedback structure. Figure \ref{fig:trajicf1} (solid red, dash magenta, dash cyan) shows adaptation is achieved for the architecture 1 ICF circuit. 
The signal strengths of the process and controller is represented by the protein concentrations. In the case of architecture 1, the signals from $p53^*$ are positive positive (transcriptional activator) while the signals from $tr$ is negative for transcriptional repressors. Therefore activation of $tr$ by the $p53^*$ proteins and repression of $p53$ by the protein expression of $tr$, meet the requirement of negative feedback structure as it effectively places the $tr$ as a controller in the negative feedback loop. The rate of expression TALE repressor ($tr$) depends on the concentration level of active p53 (p53*), each expression step functions as an integrator \cite{ang2010}. Equation \ref{eqn:eqn13} details the mRNA expression of TALE repressor activated by active p53 (p53*). Meanwhile the translation of the protein TALE repressor is governed by Equation \ref{eqn:eqn14}. The approach that we apply here assumes basal transcription rate of the TALE and using Hill functions to describe repression of its target and input activation by active p53 (p53*) protein. The enzyme catalyzed degradation is assumed to be Michaelis-menten kinetics.
\begin{align}
\frac{d[tr]}{dt} &= kcmv	+ 	k_c \left(\frac{[p53^*]^n}{{k_{mr}}^n + [p53^*]^n}\right ) - deg_{tr}.\frac{[tr]}{kmx+[tr]} \label{eqn:eqn13}\\
\frac{d[TR]}{dt} &= ktr.[tr] - degTR.[TR] \label{eqn:eqn14}
\end{align}
In order for the system to be perfectly adapting, Equation \ref{eqn:eqn13} must not be not directly dependent on either input, $u$ or $tr$. To keep the rate of change of $tr$ independent of its own concentration level, the substrate-enzyme binding affinity must be increased, effectively lowering $kmx$. In Equation \ref{eqn:eqn13}, a low $kmx$ value results in $\frac{[tr]}{kmx + [tr]} \rightarrow 1$. Hence the removal rate of the controller is of zeroth-order kinetics.

The setpoint value, $y_0$ of the system is given by the steady state solution for p53* given by
\begin{align}
p53^*_{ss} &=    {k_{mr}}.\left( \frac{deg_{tr}-kcmv}{kcmv+k_{c}-deg_{tr}} \right)^{\frac{1}{n}}, \label{eqn:p53ss}    
\end{align}
which is independent of the input, $u$, therefore meeting the requirement of perfect adaptation.

We define disease state as a constant low level when the basal level of p53 is inhibited (stepwise perturbation to the input). For a disease state, the level of p53 will fall below the wild-type (WT) level ($\sim$0.7$\mu$M) to a very low concentration ($\sim$0.1$\mu$M), a 7 fold change in its total p53 concentration shown in Figure \ref {fig:trajicf1} in solid black line. In the disease state scenario, there is a difference (steady state output error, $e$) between the set point, $y_0$ (desired p53 WT level) and the disease steady state output, $y_{out}$ (low p53 level due to perturbation). The controller (TALE repressor) receives as input the output steady state error and will generate a control action that feed to p53 (the process) to minimize this difference. This form of integral feedback control continually draws the process output toward the set point output, promoting self-regulation. TALE repressor can be thought of as effectively absorbing the input perturbation to allow p53 to maintain its wild-type steady state.  

The performance of the integral controller of architecture 1 circuit is measured by its ability to adapt and recover its wild-type concentration level when subjected with step input perturbation. It is also crucial to have minimum transient response and a reasonable rise time. We show in Figure \ref{fig:trajicf1} the results of the numerical simulation of architecture 1 circuit under a step input perturbation to the basal transcription rate of p53. The typical wild-type (no perturbation introduced) total concentration of p53 is shown in solid green line. At t=2000 minutes, a step input disturbance was subjected to the basal transcription rate of p53. The architecture 1 circuit adapts and returns the p53 level to its wild-type state (solid red) while the absence of the architecture 1 circuit shows the p53 levels attenuated to a critically low level (solid black). The recovered p53 level shown in solid red adapts to its wild-type level within the 5\% tolerance \cite{ang2013}. The nominal perturbation magnitude $u_1$ ($u_1=0.666$) is chosen such that it reflects the controller optimal performance; to adapt without overshoot with the least rise time as shown by the red line in Figure \ref{fig:trajicf1}. This value $u_1$ is a constant value ($0<u_1<1$) for architecture 1 circuit. Value of $u_1\sim 0$ reflects the strongest perturbation while $u_1 \sim 1$ represent weak perturbation. We are also interested in the scenarios where variations of the strength of perturbation are subjected to p53 transcription rate around this nominal perturbation value. We find that at 10\% stronger perturbation, the p53 system exhibit transient response before it adapts to its wild-type state (dashed magenta). At 10\% weaker perturbation, the p53 system adapts to its wild-type state albeit some delay in its rise time (dashed cyan) compared to the nominal perturbation scenario (solid red). The architecture 1 also shows that it can adapt to a wide range of input perturbations values shown in Table \ref{tab:icfperf1}. For input perturbation between $0.05<u_1<0.2$, the p53 levels show oscillatory pattern before adaptation. Meanwhile for $0.3<u_1<0.6$, the system shows some overshoot before adaptation. No oscillatory pattern is observed in this case.

\begin{table}[h]
\caption{Architecture 1 shows adaptation in the p53 levels for the widest range of perturbation values. $\dagger$ represent adaptation after oscillatory behavior of p53 levels. $\ast$ represents transient response before adaptation without oscillation. $na$ are conditions where the circuit is not adapting.}
\vspace{2mm}
\begin{tabular}{|c|c|c|c|c|c|c|c|c|c|}
\hline
\multicolumn{1}{|l|}{}                                               & \multicolumn{9}{c|}{Perturbation Strength (u)}               \\ \hline
\begin{tabular}[c]{@{}c@{}}Adaptation time \\ (minutes)\end{tabular} & 0.05 & 0.1  & 0.2  & 0.3  & 0.4  & 0.5  & 0.6  & 0.7  & 0.8  \\ \hline
Architecture 1                                                               & 4500$\dagger$ & 3300$\dagger$ & 3800$\dagger$ & 3450$\ast$ & 3500$\ast$ & 3600$\ast$ & 4200$\ast$ & 5700 & 9710 \\ \hline
Architecture 2                                                               & na   & na   & na   & na   & na   & na   & 3600$\dagger$ & 4900$\dagger$ & 6500 \\ \hline
Architecture 3                                                               & na   & na   & na   & na   & na   & na   & na   & na   & na   \\ \hline
\end{tabular}
\label{tab:icfperf1}
\end{table}

\subsection{Architecture 2: A 2-Stage Integral Control Feedback Circuit.}
\begin{figure}
\begin{center}
\includegraphics[width=1\textwidth]{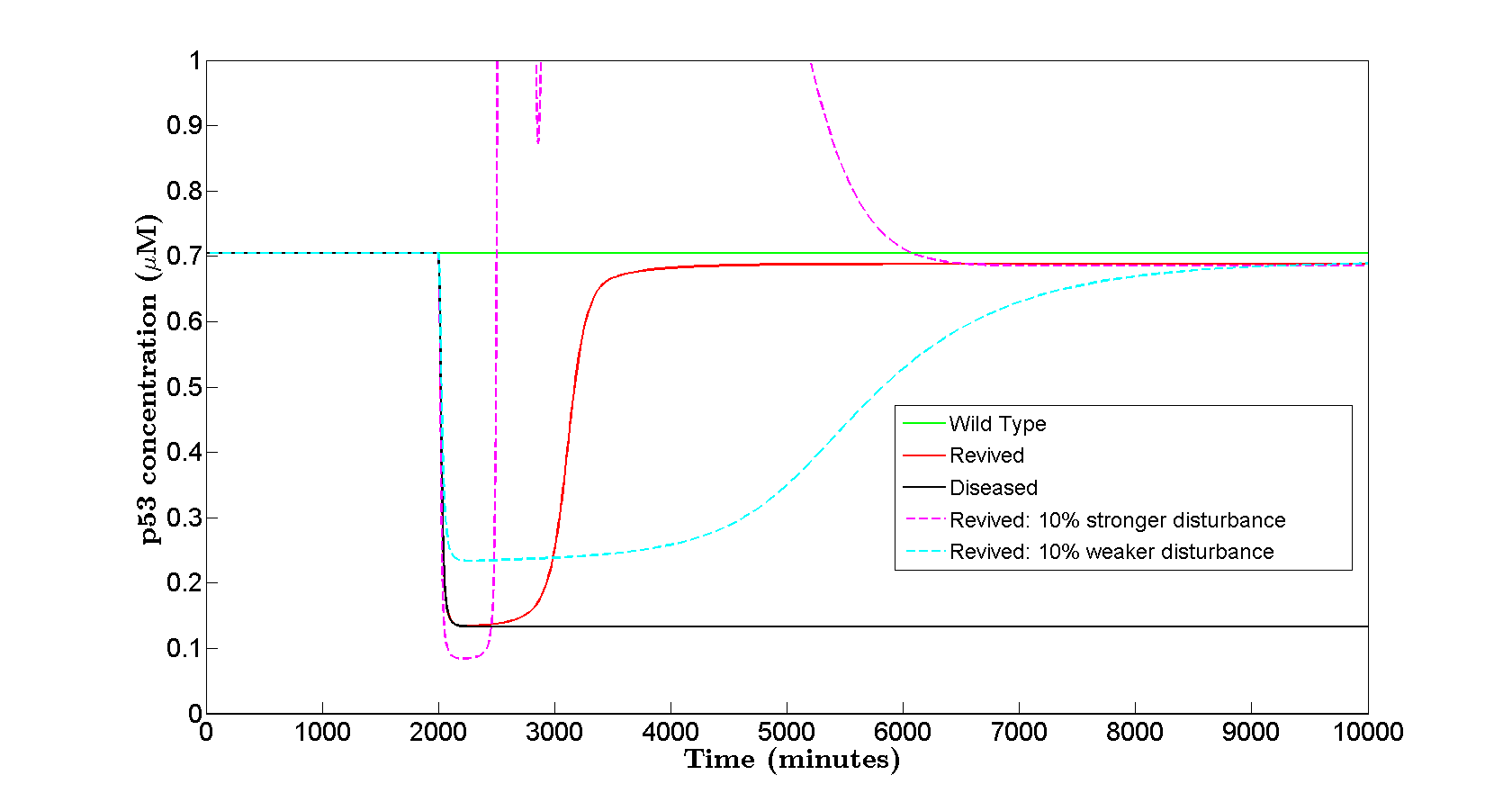}
\end{center}
\caption{Additional TALE activator introduced into the network before the controller, \textit{tr}. Time trajectory of total p53 concentration of the p53-miRNA model coupled with architecture 2 circuit. The wild-type (green) shows the typical concentration level of total p53 in a healthy cell. The diseased state (black) shows are critically low concentration level of p53 after input perturbation introduced at t=2000 minutes without the architecture 2 circuit. When the ICF (red) is introduced into the model, the p53 level return to its wild-type state. The nominal perturbation magnitude (red) is $u=0.770$. Two other scenario is also simulated; a weaker (cyan, $u=0.847$) and stronger (magenta $u=0.697$) perturbation strength was also introduced. The p53 concentration levels show adaptation where the weaker perturbation shows delayed response and the stronger perturbation shows oscillatory response.}
\label{fig:trajicf2}
\end{figure}

An additional TAL effector (TALE activator) is added to the previous single stage architecture 1 circuit, effectively creating a 2-stage integral control feedback circuit to study the effect of an additional stage to the performance of the circuit (Figure \ref{fig:icf1}(C)). This architecture 2 circuit with an additional TALE activator placed before the controller, \textit{tr} shows a certain degree in the degradation of its performance as shown in Figure \ref{fig:trajicf2}. The equation of the additional TALE activator is shown in Equation \ref{eqn:eqn15} and \ref{eqn:eqn16}. The controller, $tr$ is now activated by TALE activator and its repression target is the p53 mRNA.
\begin{align}
\frac{d[ta]}{dt} &= kcmv1	+ 	k_{c1} \left(\frac{[p53^*]^{n1}}{{k_{mr1}}^{n1} + [p53^*]^{n1}}\right ) - kdtr_{ta1}.[ta] \label{eqn:eqn15}\\
\frac{d[TA]}{dt} &= ktr.[ta] - degTA.[TA] \label{eqn:eqn16}\\
\frac{d[tr]}{dt} &= kcmv2	+ 	k_{c2} \left(\frac{[TA]^{n2}}{{k_{mr2}}^{n2} + [TA]^{n2}}\right ) - deg_{tr2}.\frac{[tr]}{kmx+[tr]} \label{eqn:eqn17}\\
\frac{d[TR]}{dt} &= ktr.[tr] - degTR.[TR] \label{eqn:eqn18}
\end{align} 

The nominal perturbation magnitude $u_2$ ($u_2=0.77$) is chosen for optimal performance for the architecture 2 circuit. At this nominal perturbation value, the system adapt without transient response with minimum rise time as shown by the red line in Figure \ref{fig:trajicf2}. The p53 level adapts to within 5\% of its wild-type p53 concentration levels. When the perturbation strength is increased by 10\% of $u_2$, oscillatory behavior appears before adaptation takes place as shown in dashed magenta line in Figure \ref{fig:trajicf2}. Similarly, when the perturbation is decreased by 10\% from $u_2$, the p53 level adapts back to its wild-type level albeit a longer rise time (dashed cyan) in Figure \ref{fig:trajicf2}. 
The range of input perturbations that this architecture can tolerate is much smaller as depicted in Table \ref{tab:icfperf1}. Much of the input range shows no adaptation ($ 0.05<u_2<0.5$). The input range between $0.6<u_2<0.7$ shows oscillatory behavior before adaptation.

\subsection{Architecture 3: Additional TALE activator transcriptionally activates p53.}
\begin{figure}
\begin{center}
\includegraphics[width=1\textwidth]{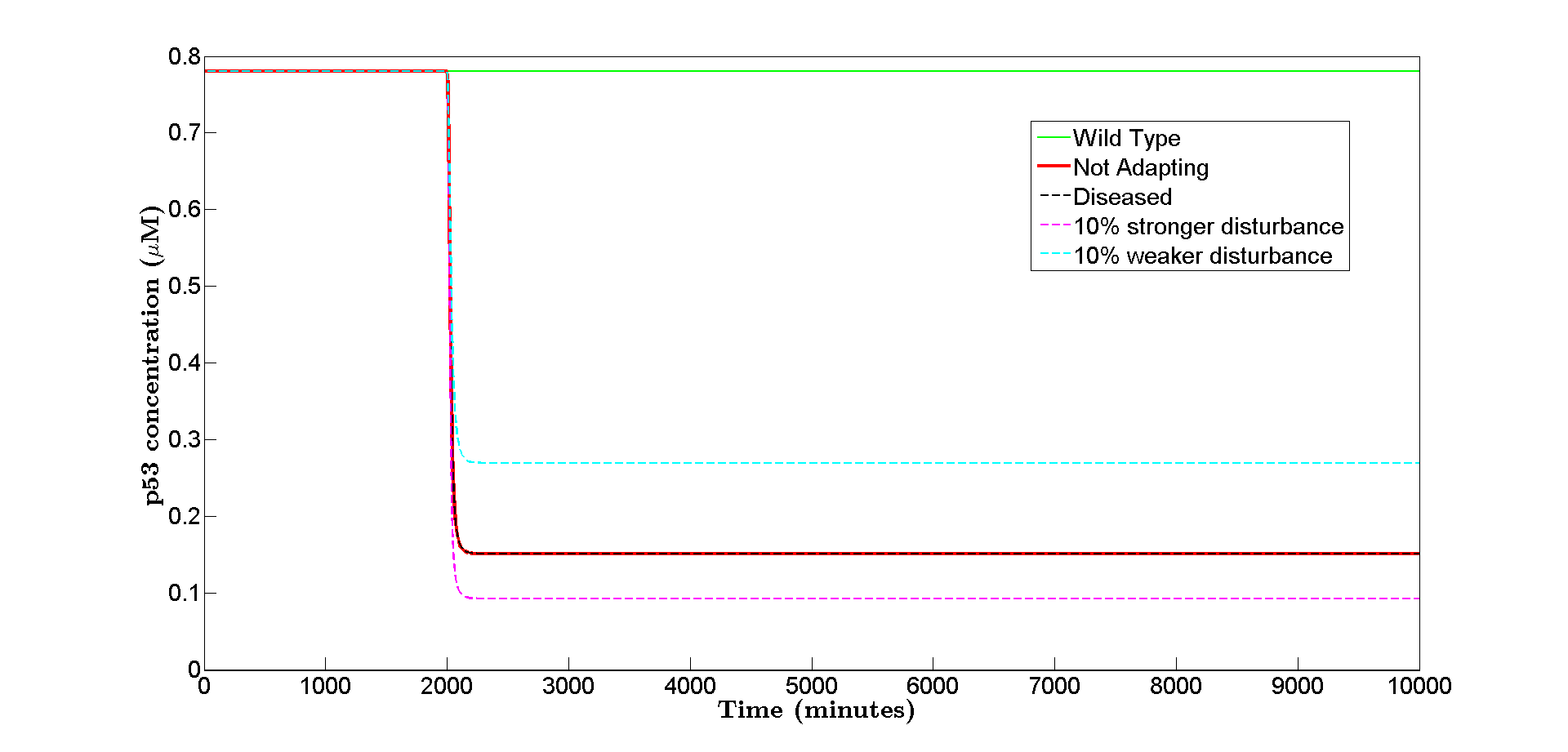}
\end{center}
\caption{Time trajectory of total p53 concentration of the p53-miRNA model coupled with ICF architecture 3 circuit. The wild-type (green) shows the typical concentration level of total p53 in a healthy cell. The diseased state (black) shows are critically low concentration level of p53 after input perturbation introduced at t=2000 minutes without the ICF architecture 3 circuit. When the ICF (red) is introduced into the model, the p53 level remains at diseased level. Two other scenario is also simulated; a weaker (cyan) and stronger (magenta) perturbation strength was also introduced. In both cases, the p53 concentration levels show no adaptation.}
\label{fig:trajicf3}
\end{figure}

We place the TALE activator, $ta$ at the last stage of the circuit to activate p53 while the controller $tr$ inhibits the action of this $ta$. The net regulatory effect of the circuit on p53 remains a negative feedback as shown in Figure \ref{fig:icf1}(D). The equation of an extra TALE activator is shown in Equation \ref{eqn:eqn21} and \ref{eqn:eqn22}. The controller, $tr$ represses the additional TALE activator which target the p53 mRNA for activation for architecture 3 given by the following equations:
\begin{align}
\frac{d[tr]}{dt} &= kcmv	+ 	k_c \left(\frac{[p53^*]^n}{{k_{mr}}^n + [p53^*]^n}\right ) - deg_{tr}.\frac{[tr]}{kmx+[tr]} \label{eqn:eqn19}\\
\frac{d[TR]}{dt} &= ktr.[tr] - degTR.[TR] \label{eqn:eqn20}\\
\frac{d[ta]}{dt} &= kcmv1	+ 	tmax \left(\frac{krep^{n2}}{{krep}^{n2} + [tr]^{n2}}\right ) - kdtr_{ta1}.[ta] \label{eqn:eqn21}\\
\frac{d[TA]}{dt} &= ktr.[ta] - degTA.[TA] \label{eqn:eqn22}
\end{align} 
The diseased state, plotted in dashed black and the attempt by the controller to revive the p53 level plotted in solid red, shows that the architecture 3 circuit is unable to adapt to its previous wild-type p53 level (Figure \ref{fig:trajicf3}). Likewise, in the cases where weaker and stronger perturbation are introduced, no adaptation was observed. All possible input perturbation range was attempted as shown in Table \ref{tab:icfperf1} without any observed adaptation.

\section{Discussion}
\label{s:res4}
\subsection{Robust integral control}
Robust perfect adaptation can be achieved when the output of process is not directly dependent on the control variable, $x$ or the perturbation, $u$. The control variable must also be independent of the value of the control action itself, to avoid an autoregulatory feedback in the controller. For architecture 1, we assume the controller adheres to a to a zeroth-order (constant) rate of removal. This essentially requires that the substrate, in this case $tr$ to be in saturation giving rise to the following criteria:\\
\indent $\ \ \ \ \  Criteria \ 1: [tr] \gg kmx$\\
In order to achieve this criteria in the design of the controller in architecture 1 circuit, the Hill constant, $kmx$ of the Michaelis-menten kinetics for enzyme catalyzed degradation term shown in Equation \ref{eqn:eqn13} must be sufficiently small so that $deg_{tr}$ is a constant so that the enzyme-catalyzed degradation occurs at saturation. The zeroth-order degradation rate is also necessary to switch off the control action when the system output is low. The control action of TALE must also be able to be driven by the output of the process (i.e. p53*) to between a state of increase ($d[tr]/dt >0$) and a state of decrease ($d[tr]/dt <0$). In addition, the controller setpoint must also be both real and positive. This two requirement give rise to the next criteria:\\
\indent $ \ \ \ \ \ Criteria \ 2:   kcmv \ < \ deg_{tr} \ < kcmv + kc $\\
An integral controller requires that a setpoint be determined. The setpoint value is determined by the steady solution of p53* as shown in Equation \ref{eqn:p53ss}. Therefore, the ideal controller setpoint $y_{p0}$ for architecture 1 is given by
\begin{align}
y_{m0} &= kmr . \left( \frac{deg_{tr}-kcmv}{kc-deg_{tr}+kcmv}   \right)^{\frac{1}{n}}  \label{eqn:eqn19}\\
y_{p0} &= \frac{ktr.y_{m0} }{degtr}\label{eqn:eqn20}
\end{align}

\begin{figure}
\begin{center}
\includegraphics[width=1\textwidth]{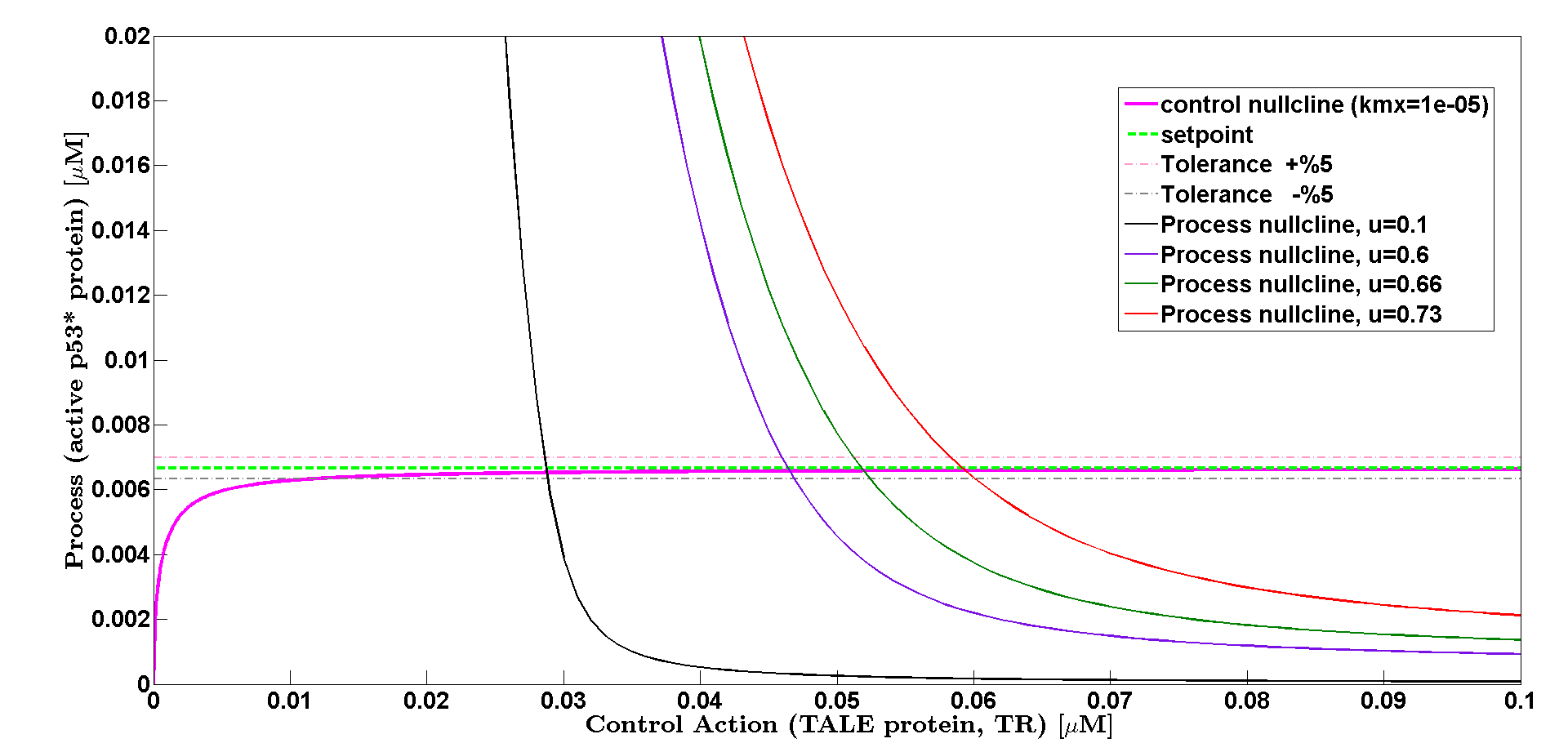}
\end{center}
\caption{The process nullclines represented by the output (active p53 (p53*)) intersect with the control action nullcline within the set tolerance (+/-5\%) of the setpoint. The nominal perturbation magnitude, $u_1=0.66$, the +/- 10\% within $u_1$ and $u=0.1$ are used to plot different process nullclines.}
\label{fig:null1}
\end{figure}

The tolerance is set to 5\% of the controller setpoint, acknowledging the fact that in practice, the adaptation to the process setpoint can only be an approximation. At saturation, $kmx$ =1\mbox{\sc{e}-}05 is sufficiently small and hence the controller nullcline shows a perfect adaptation and well within the setpoint tolerance as shown in Figure \ref{fig:null1}. Process nullcline intersect with the controller nullcline within the region of setpoint which indicates adaptation. A few perturbation magnitude, $u$ is chosen to show that at these values, adaptation is successfully achieved. For instance, at $u=0.1$ (solid black), the intersection between the controller nullcline and process nullcline takes place within the 5\% tolerance of the controller setpoint, which means perfect adaptation was achieved. Likewise, for $u=0.6$, adaptation took place as the intersection of the controller nullcline and process nullcline occurs at the controller setpoint. While, there exist transient response (minimal overshoot) or slow adaptation as detailed in Table \ref{tab:icfperf1}, nevertheless adaptation was achieved. Such minimal transient response indicated that the system error is near zero and as a result avoid the integrator windup issue.

\begin{figure}
\begin{center}
\includegraphics[width=1\textwidth]{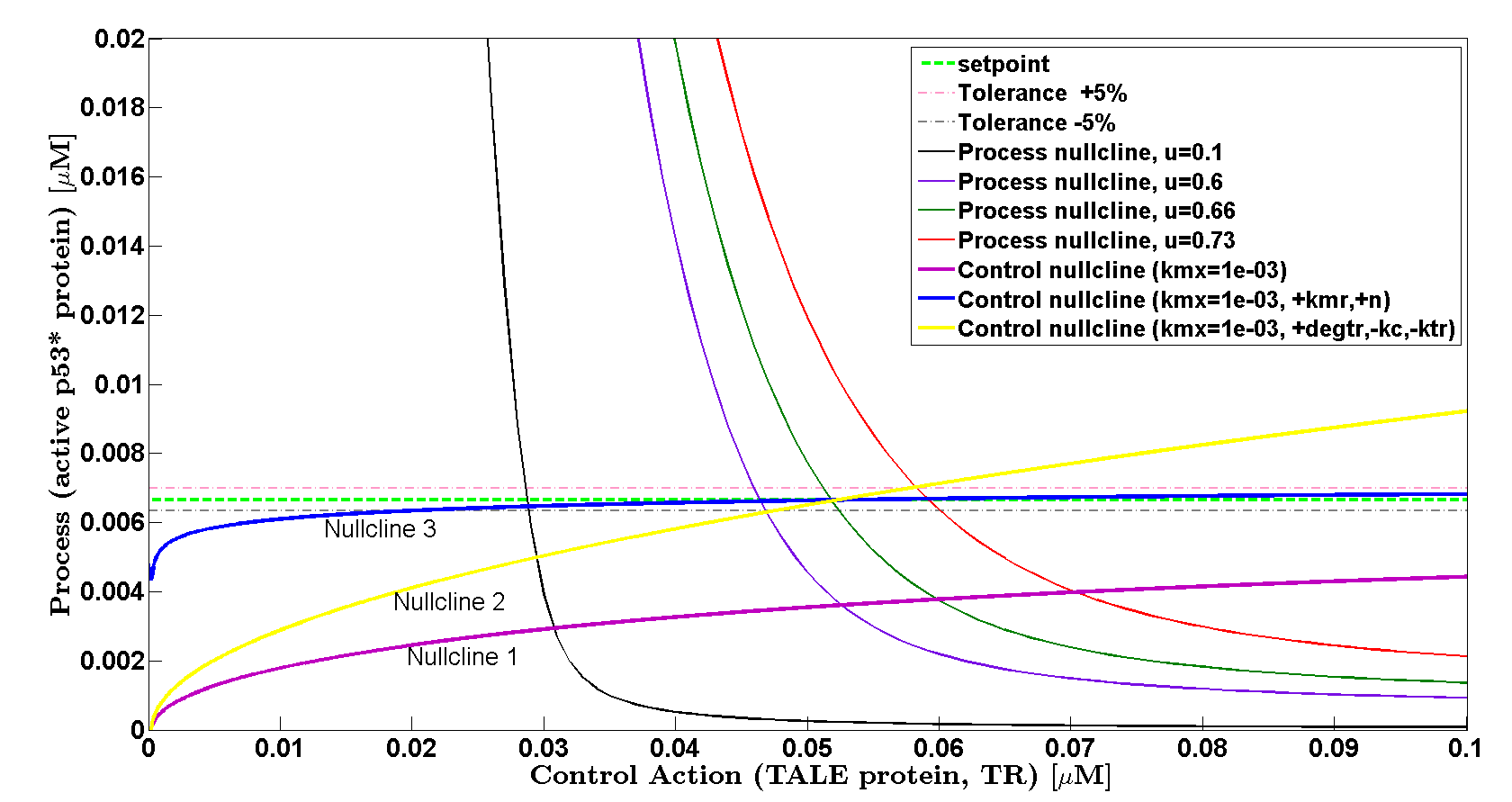}
\end{center}
\caption{Performance of the controller degrades for architecture 1 circuit as the Hill constant of the Michaelis-menten degradation term, $kmx$ is increased to 1\mbox{\sc{e}-}03 (solid magenta line). There is no intersection between the controller nullcline (nullcline 1) and the process nullcline at the setpoint. Manipulation of parameters of the controller could possible recover the function of the controller and continue to adapt as shown in the case of the nullcline 2 (solid blue line) and nullcline 3 (solid yellow line).}
\label{fig:null2}
\end{figure}

While the total amount of p53 consist of inactive p53 and active p53 (p53*), the output which feeds to the controller is the active p53 (p53*). Analytical solution of the controller can give better insights of optimal parameters for the design of the controller. Assuming that the TALE repressor (equation \ref{eqn:eqn13} holds at steady state, we can use criteria 1 and the fixed level of active p53 protein levels desired to probe relationship between parameters for the controller ($tr$) and the process ($p53$). From equation \ref{eqn:eqn13} at steady state, we obtain the analytical solution of active p53 as 
\begin{align}
[p53^*] &=k_{mr} . \left(  \frac{deg_{tr}.\biggl[\frac{[tr]}{[tr] + kmx}  \biggr] - kcmv }{k_c - deg_{tr}.\biggl[\frac{[tr]}{[tr] + kmx}  \biggr] + kcmv} \right)^{\frac{1}{n}}\label{eqn:eqn21}
\end{align} 
The TALE protein which is the control variable is represented by equation \ref{eqn:eqn14} and hence we have the final analytical solution as 
\begin{align}
[tr] &= \frac{[TR].degtr}{ktr}\label{eqn:eqn22}\\
[p53^*] &=k_{mr} . \left(  \frac{deg_{tr}.\biggl[\frac{\frac{[TR].degtr}{ktr}}{\frac{[TR].degtr}{ktr} + kmx}  \biggr] - kcmv }{k_c - deg_{tr}.\biggl[\frac{\frac{[TR].degtr}{ktr}}{\frac{[TR].degtr}{ktr} + kmx}  \biggr] + kcmv} \right)^{\frac{1}{n}}\label{eqn:eqn23}
\end{align}
In the diseased state, p53* will drop to a low level. From the analytical solution in equation \ref{eqn:eqn23}, the increased value of parameter $kmr$ and $n$ compensate for the lost in active p53 (p53*) concentration level. To simulate this case, the value of the Hill constant of the Michaelis-menten degradation term, $kmx$ is increased to $kmx$ =1\mbox{\sc{e}-}03, implying lower enzyme saturation. The controller nullcline shown in Figure \ref{fig:null2} (solid magenta) indicates that adaptation no longer takes place for the all the perturbation magnitude values simulated. To improve the performance of the controller, increasing $kmr$ and $n$ simultaneously (8 fold increase in both parameters) improve the controller performance as shown in Figure  \ref{fig:null2} (solid blue). However, increasing $kmr$ and $n$ is not practical in an experimental setting. Another approach would be to tune parameters that are practical to carry out in the lab, such as TALE protein degradation rate, $degtr$ ($degtr*1.15$) and TALE translation rate, $ktr$ ($ktr/0.56$) and maximum expression rate of the Hill function of $tr$ mRNA, $k_c$ ($k_c/1.01$). The analytical equation \ref{eqn:eqn23} shows that $degtr$ can be increased, $k_c$ and $ktr$ can be decreased to compensate for the lost in active p53 (p53*). Figure \ref{fig:null2} (solid yellow) shows that it can adapt to a smaller region of adaptation around the nominal perturbation magnitude $u=0.66$.

\begin{figure}
\begin{center}
\includegraphics[width=1\textwidth]{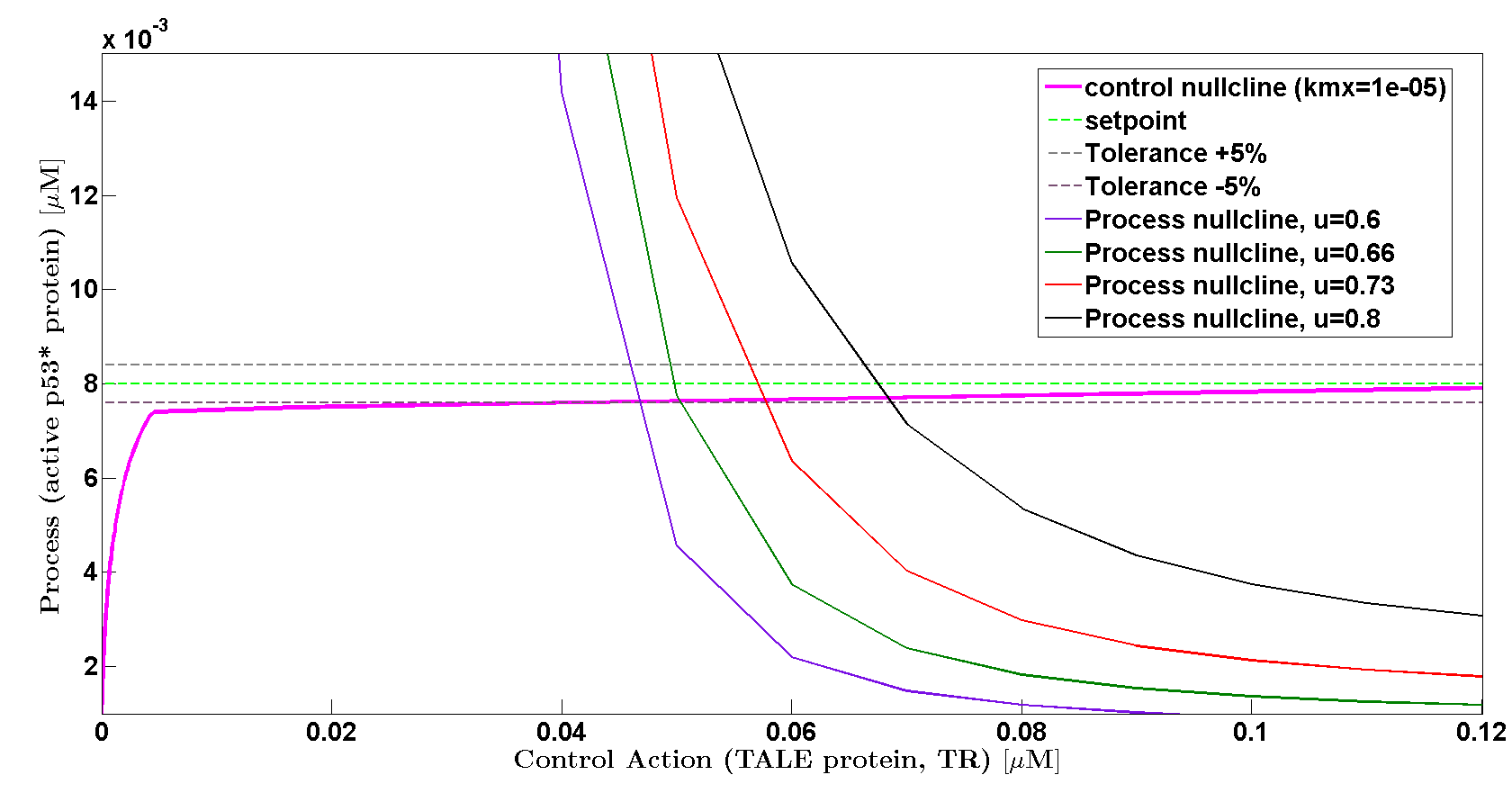}
\end{center}
\caption{Architecture 2 circuit controller nullcline intersects with setpoint tolerance -5\% at process nullcline, $u=0.6$ (purple solid line) is the maximum tolerable perturbation magnitude that the controller can adapt to. The other weaker perturbation magnitudes such as $u=0.66, 0.73, 0.8$ are also shown to adapt as the controller nullcline intersect with the process nullcline within the tolerance of the setpoint.}
\label{fig:null3}
\end{figure}

The addition of a TALE (i.e. TALE activator) before the controller (Architecture 2) limits the range of which adaptation can be achieved as detailed in Table \ref{tab:icfperf1}: Architecture 2. Adaptation was not met for all stronger perturbation magnitude $(0.05<u_2<0.5)$ and adaptation only took place for weaker perturbation magnitude $(0.6<u_2<0.8)$. The loss in the tolerable input perturbation range can be explained from the nullcline of the process and controller shown in Figure \ref{fig:null3}. The numerical approach here shows that the controller nullcline steep ascend could only intersects with the setpoint tolerance (- 5\%) at process nullcline perturbation of $u=0.6$. This agrees with the subset of the perturbation values, $0.693<u_2<0.847$ shown by the time trajectory of the simulation in Figure \ref{fig:trajicf2} and the results in Table \ref{tab:icfperf1} which includes perturbation values $0.6<u_2<0.8$. The large overshoot and long rise time of the p53* concentration for this scenario is largely attributed to the system output error which causes the control action to build up over time. This phenomenon termed as integral windup which describes how the process has reached its saturation point. In the case of Architecture 2, the process saturation is propagated to the TALE activator, $ta$ and further exacerbated the integral windup issue.

\begin{figure}
\begin{center}
\includegraphics[width=1\textwidth]{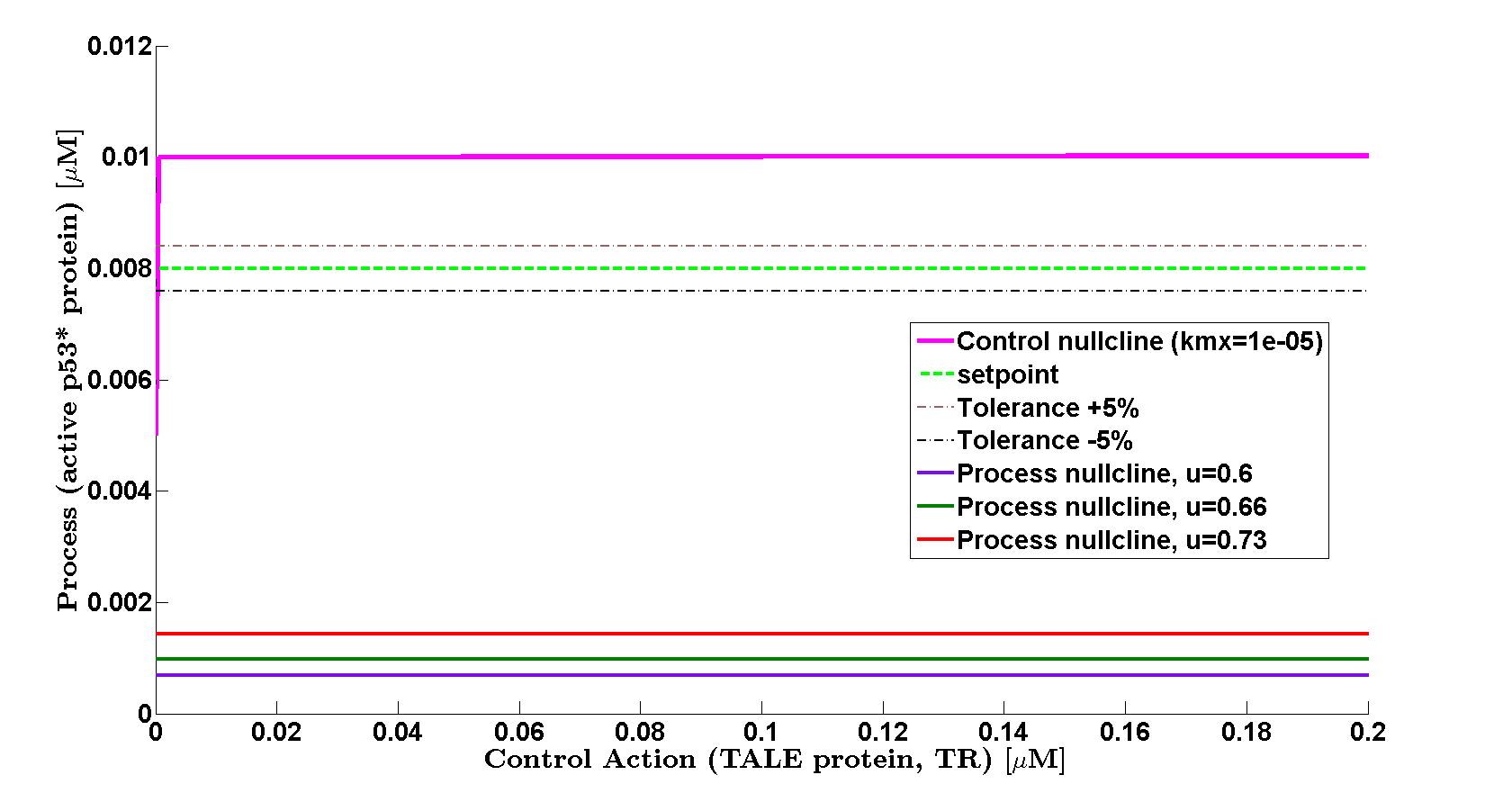}
\end{center}
\caption{Architecture 3 circuit controller nullcline shows no intersections between the controller nullcline and the process nullcline for all perturbation values.}
\label{fig:null4}
\end{figure}

Architecture 3 shows no ability for adaptation for any of the perturbation magnitude introduced. Figure \ref{fig:trajicf3} shows the results of simulation for some typical perturbation values. We hypothesize that while the integral control of the TALE repressor, $tr$ continues to output the control action, the subsequent TALE activator, $ta$ has reached its saturation point and its maximum expression level fails to activate the p53 expression effectively. The nullcline of this architecture does not show any intersection between the process and the controller (Figure \ref{fig:null4}), hence confirming the loss in the ability for the synthetic circuit of architecture 3 to adapt and return the system to it pre-stimulus level.

\section{Conclusion}
We have shown that the most basic circuit as in Architecture 1 circuit, is an effective strategy to implement the integral control feedback circuit. The parameters discussed can be optimally tuned in an experimental setting to reactivate the p53 concentration level. Additional synthetic components not only increase the complexity of rewiring the p53 systems but inflict a loss in the performance of the ICF circuit. We have demonstrated through simulations that additional components in the model have negative impact on the circuit ability to achieve adaption and the integral feedback control breaks down.

\section{Future Directions}
The robustness of the synthetic circuit architectures reported above can be further investigated. Under certain network size and parameters, the principles of integral control are no longer obeyed, suggesting certain parameter range that confer sensitivity to the model. The network design could plausibly contributes to parameter sensitivity and further investigation to the network structure could elucidate principles in which govern the robustness of network with varying sizes. 

\bibliography{mybibfile,scibib}
\bibliographystyle{Science}

\clearpage

\end{document}